\documentclass{paperClass}

\begin{document}
	
\title{KATRIN background due to surface radioimpurities}

\newcommand{\berlin}{Institut f\"{u}r Physik, Humboldt-Universit\"{a}t zu Berlin, Newtonstr. 15, 12489 Berlin, Germany}
\newcommand{\bonn}{Helmholtz-Institut f\"{u}r Strahlen- und Kernphysik, Rheinische Friedrich-Wilhelms-Universit\"{a}t Bonn, Nussallee 14-16, 53115 Bonn, Germany}
\newcommand{\cern}{CERN, Espl. des Particules 1, 1211 Meyrin, Switzerland}
\newcommand{\cmu}{Department of Physics, Carnegie Mellon University, Pittsburgh, PA 15213, USA}
\newcommand{\cwru}{Department of Physics, Case Western Reserve University, Cleveland, OH 44106, USA}
\newcommand{\etp}{Institute of Experimental Particle Physics~(ETP), Karlsruhe Institute of Technology~(KIT), Wolfgang-Gaede-Str. 1, 76131 Karlsruhe, Germany}
\newcommand{\fulda}{University of Applied Sciences~(HFD)~Fulda, Leipziger Str.~123, 36037 Fulda, Germany}
\newcommand{\laue}{Institut Laue-Langevin, 38042 Grenoble, France}
%
%
\newcommand{\ikp}{Institut f\"{u}r Astroteilchenphysik~(IAP), Karlsruhe Institute of Technology~(KIT), Hermann-von-Helmholtz-Platz 1, 76344 Eggenstein-Leopoldshafen, Germany}
\newcommand{\ipe}{Institute for Data Processing and Electronics~(IPE), Karlsruhe Institute of Technology~(KIT), Hermann-von-Helmholtz-Platz 1, 76344 Eggenstein-Leopoldshafen, Germany}
\newcommand{\itep}{Institute for Technical Physics~(ITEP), Karlsruhe Institute of Technology~(KIT), Hermann-von-Helmholtz-Platz 1, 76344 Eggenstein-Leopoldshafen, Germany}
\newcommand{\ppq}{Project, Process, and Quality Management~(PPQ), Karlsruhe Institute of Technology~(KIT), Hermann-von-Helmholtz-Platz 1, 76344 Eggenstein-Leopoldshafen, Germany    }
%
%
\newcommand{\inr}{Institute for Nuclear Research of Russian Academy of Sciences, 60th October Anniversary Prospect 7a, 117312 Moscow, Russia}
\newcommand{\lbnl}{Institute for Nuclear and Particle Astrophysics and Nuclear Science Division, Lawrence Berkeley National Laboratory, Berkeley, CA 94720, USA}
\newcommand{\madrid}{Departamento de Qu\'{i}mica F\'{i}sica Aplicada, Universidad Autonoma de Madrid, Campus de Cantoblanco, 28049 Madrid, Spain}
\newcommand{\mainz}{Institut f\"{u}r Physik, Johannes-Gutenberg-Universit\"{a}t Mainz, 55099 Mainz, Germany}
\newcommand{\mpp}{Max-Planck-Institut f\"{u}r Physik, F\"{o}hringer Ring 6, 80805 M\"{u}nchen, Germany}
\newcommand{\massit}{Laboratory for Nuclear Science, Massachusetts Institute of Technology, 77 Massachusetts Ave, Cambridge, MA 02139, USA}
\newcommand{\mpik}{Max-Planck-Institut f\"{u}r Kernphysik, Saupfercheckweg 1, 69117 Heidelberg, Germany}
\newcommand{\muenster}{Institut f\"{u}r Kernphysik, Westf\"alische Wilhelms-Universit\"{a}t M\"{u}nster, Wilhelm-Klemm-Str. 9, 48149 M\"{u}nster, Germany}
\newcommand{\npi}{Nuclear Physics Institute of the CAS, v. v. i., CZ-250 68 \v{R}e\v{z}, Czech Republic}
\newcommand{\unc}{Department of Physics and Astronomy, University of North Carolina, Chapel Hill, NC 27599, USA}
\newcommand{\washington}{Center for Experimental Nuclear Physics and Astrophysics, and Dept.~of Physics, University of Washington, Seattle, WA 98195, USA}
\newcommand{\wuppertal}{Department of Physics, Faculty of Mathematics and Natural Sciences, University of Wuppertal, Gau{\ss}str. 20, 42119 Wuppertal, Germany}
\newcommand{\saclay}{IRFU (DPhP \& APC), CEA, Universit\'{e} Paris-Saclay, 91191 Gif-sur-Yvette, France }
\newcommand{\tum}{Technische Universit\"{a}t M\"{u}nchen, James-Franck-Str. 1, 85748 Garching, Germany}
\newcommand{\tunl}{Triangle Universities Nuclear Laboratory, Durham, NC 27708, USA}
%
%
\newcommand{\ornl}{Also affiliated with Oak Ridge National Laboratory, Oak Ridge, TN 37831, USA}
%
%
%
\author{Florian Fränkle}
\affiliation{\ikp{}}
\author{Anna Schaller}
\affiliation{\tum{}}
\affiliation{\mpp{}}
\author{Christian Weinheimer}
\affiliation{\muenster{}}
\author{Guido Drexlin}
\affiliation{\etp{}}
\author{Susanne Mertens}
\affiliation{\tum{}}
\affiliation{\mpp{}}
\author{Klaus Blaum}
\affiliation{\mpik{}}
\author{Ernst Otten}
\altaffiliation{Deceased}
\affiliation{\mainz{}}
\author{Volker Hannen}
\affiliation{\muenster{}}
\author{Lutz Bornschein}
\affiliation{\ikp{}}
\author{Joachim Wolf}
\affiliation{\etp{}}
\author{Klaus Schlösser}
\affiliation{\ikp{}}
\author{Frank Müller}
\affiliation{\mpik{}}
\author{Thomas Thümmler}
\affiliation{\ikp{}}
\author{Ferenc Glück}
\affiliation{\ikp{}}
\author{Alexander Osipowicz}
\affiliation{\fulda{}}
\author{Dominic Hinz}
\affiliation{\ikp{}}
\author{Fabian Harms}
\affiliation{\etp{}}
\author{Philipp Ranitzsch}
\affiliation{\muenster{}}
\author{Nikolaus Trost}
\affiliation{\ikp{}}
\author{Jonas Karthein}
\affiliation{\cern{}}
\affiliation{\mpik{}}
\author{Ulli Köster}
\affiliation{\laue{}}
\author{Karl Johnston}
\affiliation{\cern{}}
\author{Alexey Lokhov}
\affiliation{\muenster{}}

\date{\today}

\begin{abstract}
The goal of the KArlsruhe TRItrium Neutrino (KATRIN) experiment is the determination of the effective electron antineutrino mass with a sensitivity of \SI{0.2}{\electronvolt/c^2} at \SI{90}{\percent} C.L. \footnote{C.L. - confidence level}. 
This goal can only be achieved with a very low background level in the order of \SI{10}{\milli\cps}\footnote{mcps - milli count per second}. 
A possible background source is $\alpha$-decays on the inner surface of the KATRIN Main Spectrometer.
Two $\alpha$-sources, \radium{223} and \thorium{228},  were installed at the Main Spectrometer with the purpose of temporarily increasing the background in order to study $\alpha$-decay induced background processes.
In this paper, we present a possible background generation mechanism and measurements performed with these two radioactive sources. 
Our results show a clear correlation between $\alpha$-activity on the inner spectrometer surface and  background from the volume of the spectrometer. 
Two key characteristics of the Main Spectrometer background --- the dependency on the inner electrode offset potential, and the radial distribution --- could be reproduced with this artificially induced background. These findings indicate a high contribution of $\alpha$-decay induced events to the residual KATRIN background.
\end{abstract}

\maketitle


\section{Introduction}
\label{sec:introduction}

The KArlsruhe TRItrium Neutrino (KATRIN) experiment aims to improve present neutrino mass limits by one order of magnitude to \SI{0.2}{\electronvolt/c^2} at \SI{90}{\percent} C.L.~through the investigation of the tritium $\upbeta$-electron energy spectrum close to its endpoint \cite{KATRINCollaboration2005, Drexlin2013}. 
This approach is based on the kinematics of the $\upbeta$-decay and is therefore model-independent. The observable of the experiment is the incoherent sum of the three mass eigenstates $m_{i}$ and the neutrino matrix element $U_{ei}$:

\begin{align*}
	m_{\beta}^{2} = \sum_{i}|U_{ei}|^{2}m_{i}^{2}.
\end{align*}
	
\begin{figure}
	\centering
	\includegraphics[width=1.0\columnwidth]{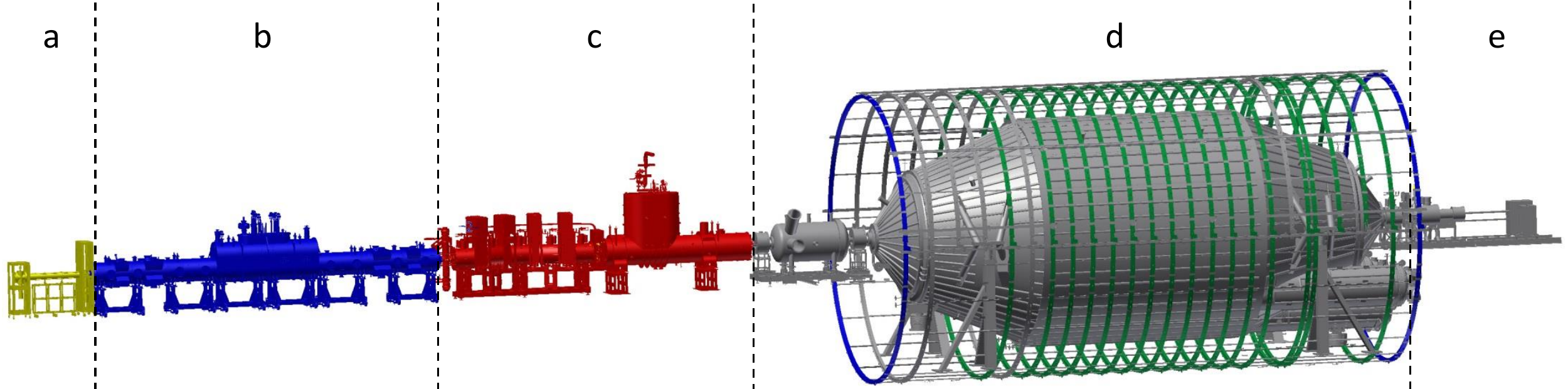}
	\caption{Scheme of the \SI{70}{\meter} long KATRIN beamline: a) rear section, b) windowless gaseous tritium source, c) transport section, d) spectrometer section with aircoil, e) detector section.}
	\label{fig:beamline}
\end{figure}

The setup of the KATRIN experiment is illustrated in \cref{fig:beamline}. For detailed specifications, the reader is referred to \cite{KATRINCollaboration2005, KATRIN2020}.
Molecular tritium is continuously injected into the windowless gaseous tritium source WGTS (b) \cite{Gehring2008}. One end of the WGTS is closed by the rear wall, located in the rear section (a). The $\upbeta$-electrons created in the WGTS are guided magnetically by superconducting magnets \cite{Arenz_2018} through the transport (c) and spectrometer section (d) and further to the detector section (e).
The tritium flow is reduced in the transport section to prevent tritium from entering the spectrometer section, where it would be a source of background. 
The spectrometer section consists of two spectrometers, the Pre- and Main Spectrometer. The Pre-Spectrometer acts as a filter for low-energy $\upbeta$-electrons. 
In the Main Spectrometer, the energies of the remaining electrons are analyzed using the principle of magnetic adiabatic collimation with an electrostatic filter (MAC-E filter) \cite{Kruit_1983, LOBASHEV1985496, picard}. The high voltage applied to the \SI{23.2}{\meter} long stainless steel vessel, fine-shaped by \num{23400} thin wires of inner electrode (IE) system, defines the filter potential $U_{\mathrm{ret}}=U_{\mathrm{vessel}}+U_{\mathrm{IE}}$. For a retarding potential of $U_{\mathrm{ret}}=$~\SI{-18.6}{\kilo\volt}, these values are typically $U_{\mathrm{IE}}=$~\SI{-0.2}{\kilo\volt} and $U_{\mathrm{vessel}}=$~\SI{-18.4}{\kilo\volt}. 
The magnetic field inside the Main Spectrometer is shaped by a \SI{12.6}{\meter} diameter air coil system \cite{Erhard_2018}, represented in \cref{fig:beamline}.
Electrons with energies above the filter potential are transmitted through the Main Spectrometer in an ultra-high vacuum of about \SI{e-11}{\milli\bar} to the detector section. Here they are post-accelerated at a dedicated Post Acceleration Electrode (PAE) with a voltage of $U_{\mathrm{PAE}}=$~\SI{10}{\kilo\volt} and counted by the focal plane detector (FPD) \cite{Amsbaugh2015} at the downstream end of the beamline. 

For the targeted neutrino mass sensitivity, a low background in the order of \SI{10}{\milli\cps} (KATRIN technical design) is required \cite{KATRINCollaboration2005}. Past investigations have shown that there is no significant contribution from muons \cite{ALTENMULLER201940}, external gamma radiation \cite{Altenmueller2019}, radon \cite{Harms2015_1000050027, goerhardt}, and Penning discharges \cite{Aker2019} during normal operation (neutrino mass measurement configuration) with active and passive countermeasures. 
One possible background source is $\alpha$-decays in the spectrometer walls, presumably from \polonium{210}-to-\lead{206} decay.  
Its origin and the associated background creation mechanism via Rydberg atoms are described in \cref{sec:rydberg}.
To investigate whether the proposed background mechanism contributes to the Main Spectrometer background, we aim to induce this background process in a controlled manner.
Two different $\alpha$-sources, \radium{223} and \thorium{228}, were installed at the KATRIN Main Spectrometer for this purpose. In both decay chains, there are no long-lived daughters that could cause long-term contamination. 
These isotopes decay via four (\radium{223}) or five (\thorium{228}) $\alpha$-decays, see \cref{equ:rachain,equ:thchain}  for the principal decay paths \cite{decaychain}, possibly inducing a large Rydberg-induced background through a series of consecutive implantations and decays.


\begin{align}
	\begin{split}
		\textsuperscript{228}\mathrm{Th}
		\xrightarrow[\text{\quad\SI{1.9}{\year}}\quad]{\alpha} \textsuperscript{224}\mathrm{Ra}
		\xrightarrow[\text{\quad\SI{3.7}{\day}}\quad]{\alpha} \textbf{\textsuperscript{220}Rn} 
		\xrightarrow[\text{\quad\SI{56}{\second}}\quad]{\alpha} \textsuperscript{216}\mathrm{Po}
		\xrightarrow[\text{\quad\SI{0.15}{\second}}\quad]{\alpha} \textbf{\textsuperscript{212}Pb}
		\\
		\textbf{\textsuperscript{212}Pb}\xrightarrow[\text{\quad\SI{10.6}{\hour}}\quad]{\beta} \textsuperscript{212}\mathrm{Bi}
		 \quad & 64\% 
		\xrightarrow[\text{\quad\SI{61}{\minute}}\quad]{\beta} \textsuperscript{212}\mathrm{Po}    
		\xrightarrow[\text{\quad\SI{310}{\nano\second}}\quad]{\alpha} \textsuperscript{208}\mathrm{Pb}	\\	
		& 36\%
		\xrightarrow[\text{\quad\SI{61}{\minute}}\quad]{\alpha} \textsuperscript{208}\mathrm{Tl}
		\xrightarrow[\text{\quad\SI{3.1}{\minute}}\quad]{\beta} \textsuperscript{208}\mathrm{Pb}  
	\end{split}
	\label{equ:thchain}
\end{align}

\begin{align}
	\textsuperscript{223}\mathrm{Ra}
	\xrightarrow[\text{\quad\SI{11}{\day}}\quad]{\alpha} \textbf{\textsuperscript{219}Rn}
	\xrightarrow[\text{\quad\SI{4}{\second}}\quad]{\alpha} \textsuperscript{215}\mathrm{Po}
	\xrightarrow[\text{\quad\SI{1.8}{\milli\second}}\quad]{\alpha} \textbf{\textsuperscript{211}Pb}
	\xrightarrow[\text{\quad\SI{36}{\minute}}\quad]{\beta} \textsuperscript{211}\mathrm{Bi}
	\xrightarrow[\text{\quad\SI{2.1}{\minute}}\quad]{\alpha} \textsuperscript{207}\mathrm{Tl} 
	\xrightarrow[\text{\quad\SI{4.8}{\minute}}\quad]{\beta} \textsuperscript{207}\mathrm{Pb}
\label{equ:rachain}  
\end{align}

The production and installation of the two radioactive sources are described in \cref{sec:production} and the performed measurements in \cref{sec:measurement}. 
We conclude our results in \cref{sec:conclusion}.

\section{The Rydberg background hypothesis}
\label{sec:rydberg}

A possible background source in KATRIN is the $\alpha$-decay of \polonium{210} to \lead{206} on the Main Spectrometer surface \cite{FraenkleProc}.
The polonium isotope is a progeny following the \lead{210} $\rightarrow$ $^{210}$Bi $\beta$-decay. The lead isotope (half-life = \SI{22.3}{\year}), a progeny of the $^{222}$Rn-decay, was most likely implanted by $\alpha$-decay during the 5-year installation period of the inner electrode system, during which the spectrometer was at ambient pressure with continuous ventilation through a HEPA filter. For low-level background experiments, \lead{210} and its progenies are a known source of background on surfaces \cite{Leung} that had been exposed to ambient air and the local concentration of $^{222}$Rn for an extended period of time. The total amount of \lead{210} accumulated on the surfaces of the Main Spectrometer is estimated to be of order kBq \cite{Harms2015_1000050027}. 

The high nuclear recoil energy of the daughter atom \lead{206} in a \polonium{210}  decay can sputter off atoms from the spectrometer surface \cite{Trost2019_1000090450}. 
The released energy also induces highly excited states, so-called Rydberg states, with ionization energies in the meV to eV range.
Emitted Rydberg atoms can be in different charge states: positive, negative, or neutral.
Charged atoms will be pulled either towards the inner electrodes, which are more negative than the wall, or back to the spectrometer wall.
Neutral Rydberg atoms propagate into the spectrometer volume and are relevant for background generation.
The atomic species of Rydberg atoms depend on the material composition of the stainless steel (AISI 316 LN \cite{KATRINCollaboration2005}) of the main spectrometer vessel. 
The primary elements are iron (\SI{65}{\percent}), chromium (\SI{17}{\percent}) and nickel (\SI{13}{\percent}). 
In addition, the passivation layer of the electropolished surface contains a high concentration of oxygen and carbon, as well as surface adsorbed hydrogen atoms, which are dominantly sputtered \cite{Lowery}. 
The accumulating hydrogen surface coverage increases the background rate as verified by measurements after bake-outs of the Main Spectrometer \cite{Harms2015_1000050027}.

Rydberg atoms leaving the surface are strongly affected by the black body radiation of the spectrometer walls at room temperature. 
Those photons stimulate photo-emission of the excited atoms, de-excitation, excitation to higher levels, and the subsequent ionization. 
Low-energy background electrons are generated if the Rydberg atoms are ionized within the spectrometer volume.
Such a mechanism via neutral Rydberg atoms is required because only electrons created inside the  magnetic flux tube mapped onto the detector can generate background events at normal magnetic field settings (neutrino mass measurement configuration).
Electrons originating from the spectrometer walls are magnetically shielded or reflected by the electric potential of the inner electrodes \cite{valerius2010}.
Subsequent implantation of daughter isotopes on the opposite surfaces of the Main Spectrometer gives rise to additional sputtering, which boosts the proposed background creation mechanism.

In addition to black body radiation, external electric or magnetic fields influence excited atoms through the Stark effect or the Zeeman effect \cite{Gallagher}.
In the KATRIN Main Spectrometer, the perturbation of the excited atoms by magnetic fields can be neglected, but not the electric field between the vessel and the wire electrode.
In this high electric field region, Rydberg atoms can be ionized.
Ionization of Rydberg atoms by collisions with residual gas can be neglected in the ultra-high vacuum of the spectrometers.
Therefore, ionization through black body radiation is considered to be the main ionization mechanism in the sensitive magnetic flux volume.

\begin{figure}
	\centering
	\includegraphics[width=0.5\columnwidth]{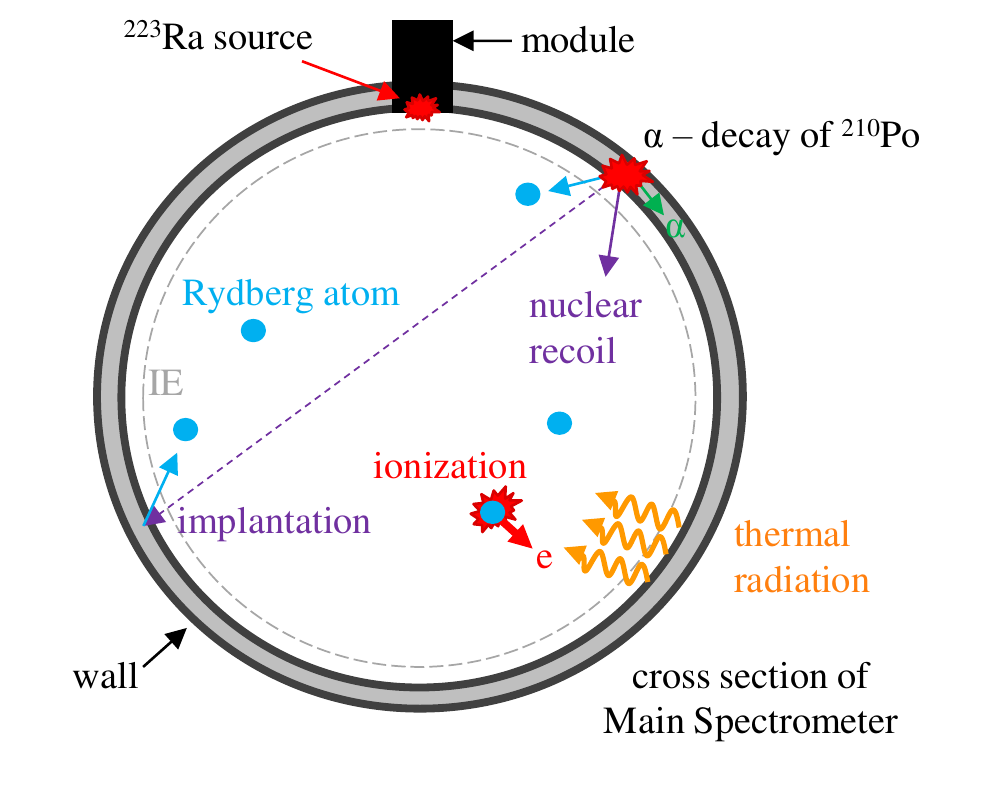}
	\caption{Schematic of the Rydberg induced background mechanism. Highly excited Rydberg atoms can be created via $\alpha$-decays in the surface and accompanying sputtering processes. If they are ionized, e.g. by thermal radiation, low-energy electrons are generated. The whole process is enhanced by the implantation of daughter isotopes which in turn initiate additional sputtering. To simulate background mechanisms such as the Rydberg mechanism, a \radium{223} source was installed at the spectrometer surface level. The wall of the vacuum vessel is at high voltage. The inner electrode (IE) is up to \SI{200}{\volt} more negative in electric potential.}
	\label{fig:rydberg}
\end{figure}

An overview of the proposed background generation process through excited neutral atoms emitted from the surface and their subsequent ionization is illustrated in \cref{fig:rydberg}. 

\section{Production and Installation of the $\alpha$-sources at the KATRIN Main Spectrometer}
\label{sec:production}

\begin{figure}
	\centering
	\includegraphics[width=0.6\columnwidth]{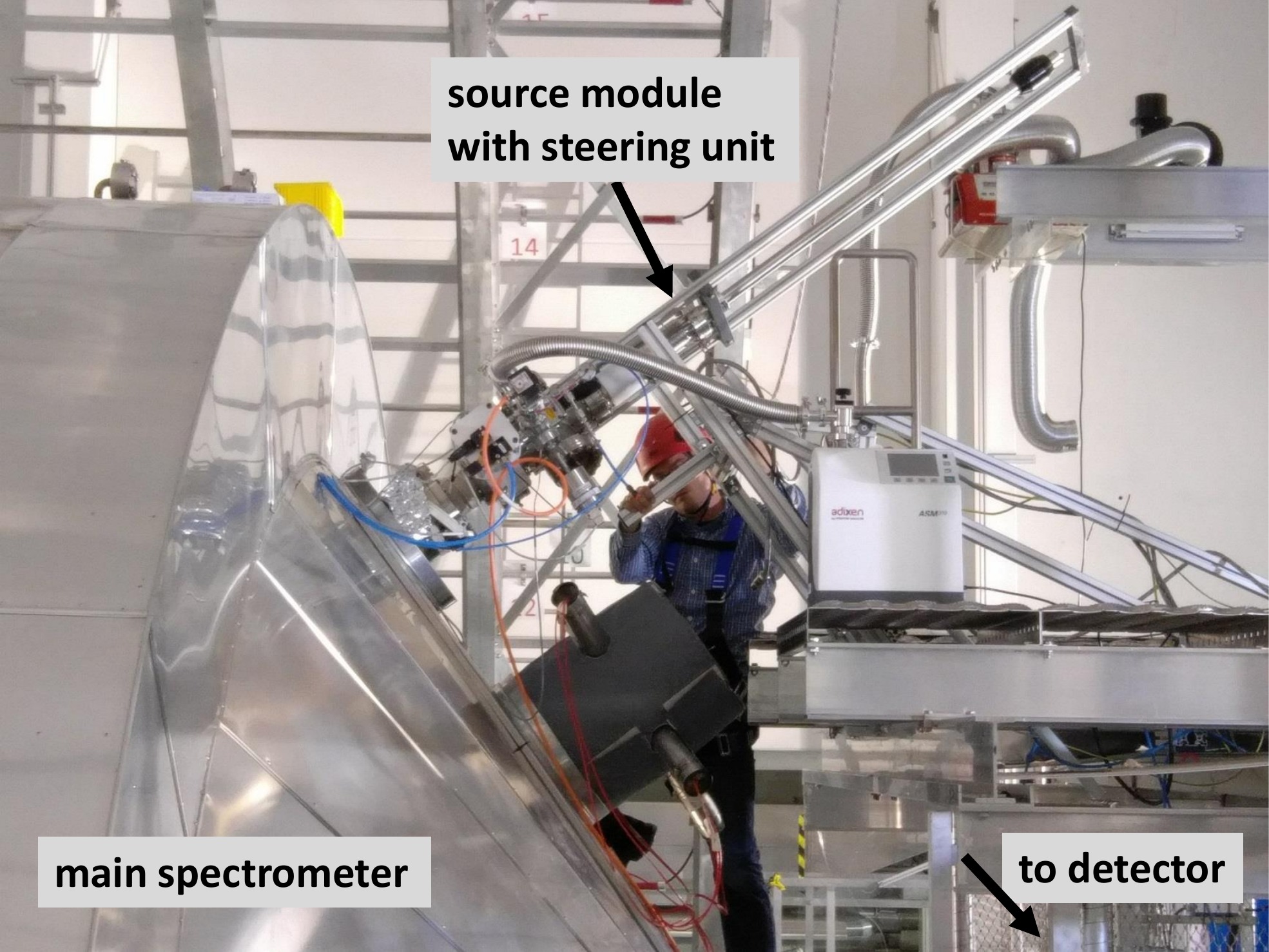}
	\caption{Installation of the \radium{223}-source on the top side of the KATRIN Main Spectrometer. The source was mounted inside the vacuum chamber at the tip of the magnetically coupled stainless steel rod of the linear motion device, which can be lowered into the spectrometer.}
	\label{fig:installation}
\end{figure}

The \thorium{228} source \cite{Lang2016} was produced by electroplating thorium nitrate onto a stainless steel disk of \SI{30}{\milli\metre} diameter. The activity of the source at the time of production in March 2015 was \SI{40}{\kilo\becquerel}. At the KATRIN Main Spectrometer, the source is installed at the same vacuum flange (Fig. \ref{fig:installation}) as the radium source.

The \radium{223} was implanted with a depth of $\approx$~\SI{40}{\angstrom} into the round side of a half-sphere gold substrate with a diameter of \SI{1}{\centi\meter}, at ISOLDE at CERN \cite{Borge2017}. 
The isotope was produced off-line from in-target decay of \thorium{227} in a UC\textsubscript{x} target that had previously been irradiated with \SI{1.4}{\giga\electronvolt} protons. The \radium{223} was thermally released from the $\approx$~\SI{2000}{\celsius} hot target and surface ionized on a $\approx$~\SI{2100}{\celsius} hot tantalum ionizer, accelerated to \SI{20}{\kilo\electronvolt}, mass separated, and implanted into the curved side of the half-sphere. 
The off-line operation was optimized to minimize contamination from the `wrong' masses in the molecular beams or other artifacts stemming from short-lived isotopes.
By the time of the measurements at KATRIN, the source had an activity of $\approx$~\SI{6}{\kilo\becquerel}. 
To introduce the source to the Main Spectrometer without breaking the ultra-high vacuum, it is mounted to a sample holder on a magnetically coupled linear-motion device in a vacuum chamber, which is connected through a gate valve to a DN200CF flange on top of the Main Spectrometer.
The magnetically coupled linear UHV flange enables the positioning of the source at the Main Spectrometer surface level, as illustrated in \cref{fig:rydberg}.
A picture of the setup is shown in \cref{fig:installation}. 

\section{Measurements with artificial background sources}
\label{sec:measurement}

\subsection{Measurement configuration}

\begin{figure*}
	\centering
	\subfigure[~volume magnetic field configuration]{%
		\includegraphics[width=0.4\columnwidth]{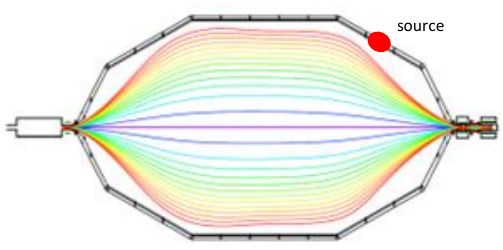}
		\label{fig:volume}
	}%
	\subfigure[~surface magnetic field configuration]{%
		\includegraphics[width=0.4\columnwidth]{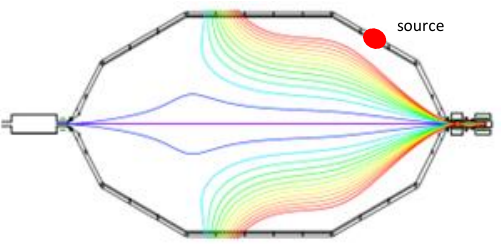}
		\label{fig:surface}
	}%
	\caption{Measurement configuration: volume \protect\subref{fig:volume} and surface \protect\subref{fig:surface} magnetic field configuration enable the investigation of volume and surface-induced events. The colored lines show the magnetic field lines reaching the silicon detector. Each line corresponds to a boundary between the ring-wise segmentation of the detector.}
	\label{fig:fieldconfig}
\end{figure*}

Two magnetic field configurations were used for the presented measurements.
Background originating from the spectrometer volume is observed in the volume magnetic field configuration as shown in \cref{fig:volume}.
To investigate the background electrons from the inner surface of the Main Spectrometer, the so-called surface magnetic field configuration is used, where the magnetic field lines point towards the cylindrical part of the spectrometer walls, as illustrated in \cref{fig:surface}. 
In the surface magnetic field configuration the magnetic flux tube is neither electrically nor magnetically shielded, leading to a much higher background rate compared to the one observed in the volume magnetic field configuration. 
To switch between these configurations the air coil currents are changed, which takes a few minutes.

The volume and surface magnetic field configurations used for the measurements with thorium are similar to the ones used with radium (shown in \cref{fig:fieldconfig}).
Slight deviations in the magnetic field configurations arise because they are realized by different air coil currents. 
The differences in the configurations affect the volume of the magnetic flux tube (volume magnetic field configuration) and the surface area that is mapped onto the detector (surface magnetic field configuration). 
Therefore, different absolute background rates are observed for the thorium and radium measurements. 

\subsection{Thorium source}

The goal of this measurement is to accumulate \lead{212} on the inner surface of the KATRIN Main Spectrometer and to observe a time-dependent background component, which is expected to decrease with the \SI{10.64 +- 0.01}{\hour} half-life \cite{Browne2005} of the isotope. 
For this purpose, the \thorium{228}-source was installed in December 2016 at the Main Spectrometer. The valve to the Main Spectrometer was opened for \SI{20}{\hour} \SI{6}{\minute} to release the emanating \radon{220} into the Main Spectrometer. The \radon{220}, which was not pumped from the spectrometer volume, subsequently decayed into \lead{212} which was implanted into the spectrometer walls.

A measurement with the two alternating magnetic field settings (see figure \ref{fig:fieldconfig}) was started shortly after the valve to the \thorium{228}-source was closed.  
Each measurement point in the volume magnetic field configuration lasted for \SI{2000}{\second} and each point in the surface magnetic field configuration lasted \SI{200}{\second}. 

\begin{figure*}
	\centering
	\subfigure[~volume magnetic field configuration]{%
		\includegraphics[width=0.4\columnwidth]{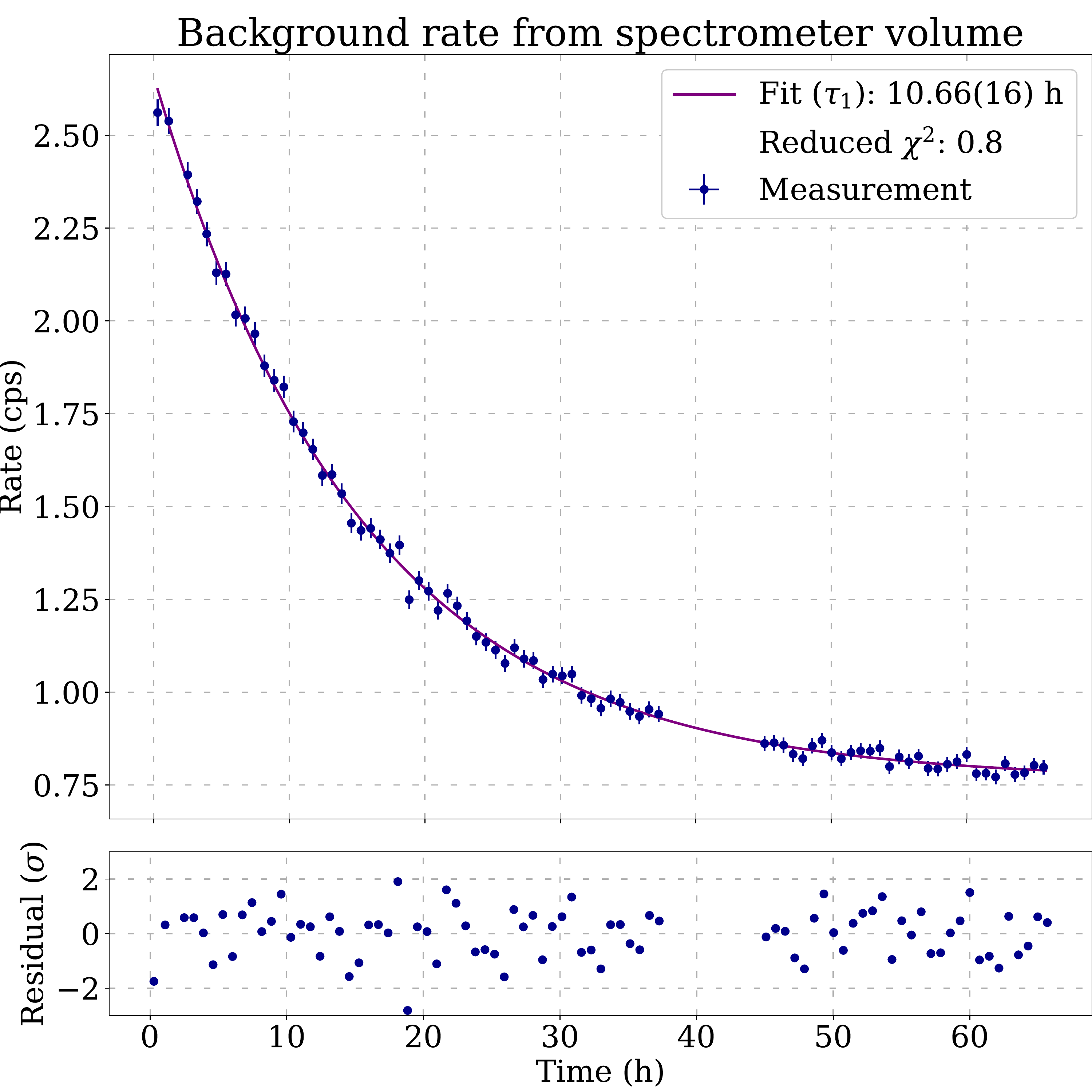}
		\label{fig:BGafter228ThVolume}
	}%
	\subfigure[~surface magnetic field configuration]{%
		\includegraphics[width=0.4\columnwidth]{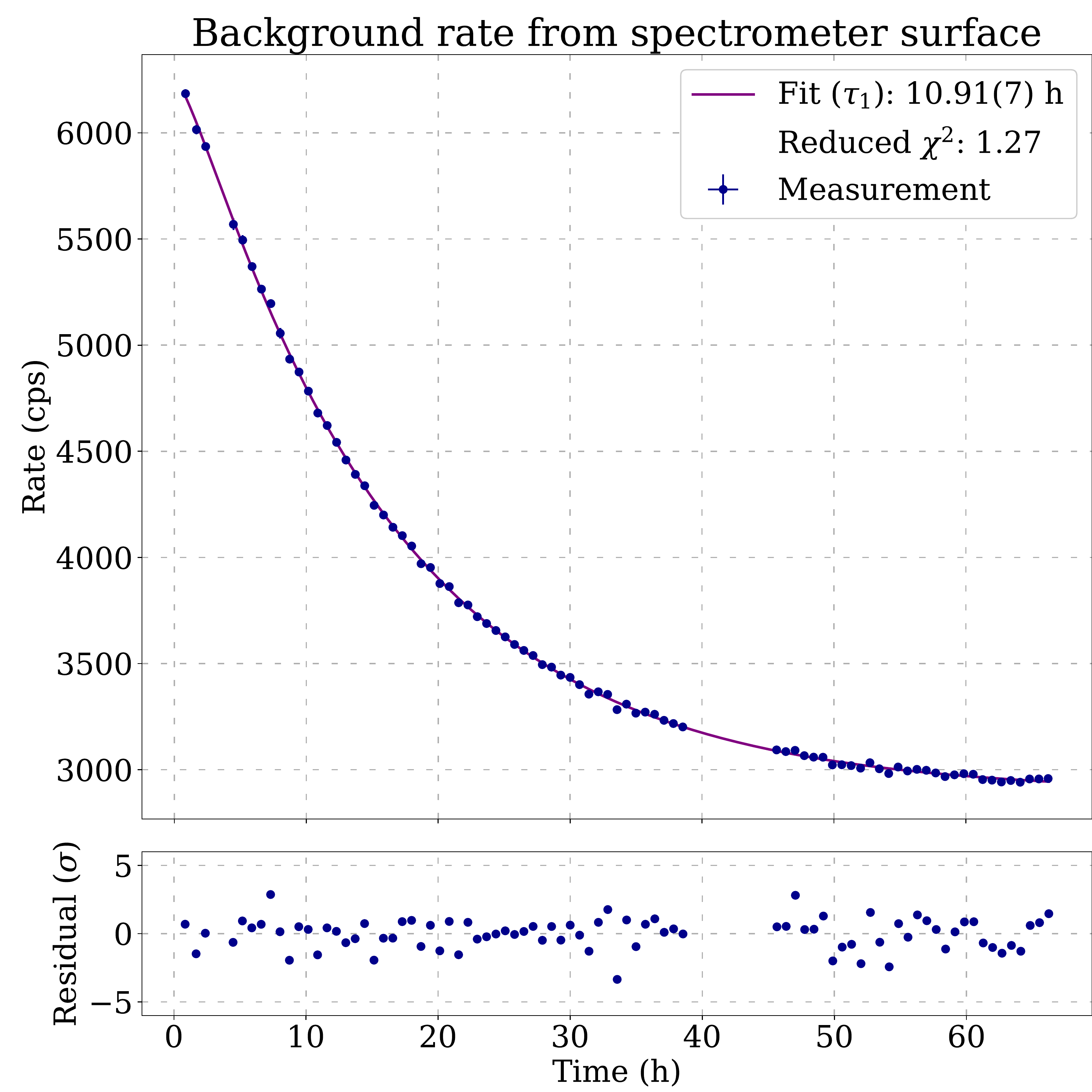}
		\label{fig:BGafter228ThSurface}
	}%
	\caption{The \thorium{228}-induced rate in volume magnetic field setting versus time after closing the valve to the \thorium{228} source \protect\subref{fig:BGafter228ThVolume}. A fit of equation \ref{eq:thoriumFitExp} to the measurement data was performed. The observed half-life of \SI{10.66 +- 0.16}{\hour} matches the literature value \SI{10.64 +- 0.01}{\hour} \cite{Browne2005} for \lead{212}. The background offset is \SI{0.763 +- 0.005}{\cps}. \thorium{218}-induced rate in surface magnetic field setting versus time after closing the valve to \thorium{228} source \protect\subref{fig:BGafter228ThSurface}. A fit of equation \ref{eq:thoriumFitBateman} to the measurement data was performed. The background offset is \SI{2891 +- 3}{\cps}.}
	\label{fig:fieldconfig}
\end{figure*}

\begin{figure}
	\centering
	\includegraphics[width=0.5\columnwidth]{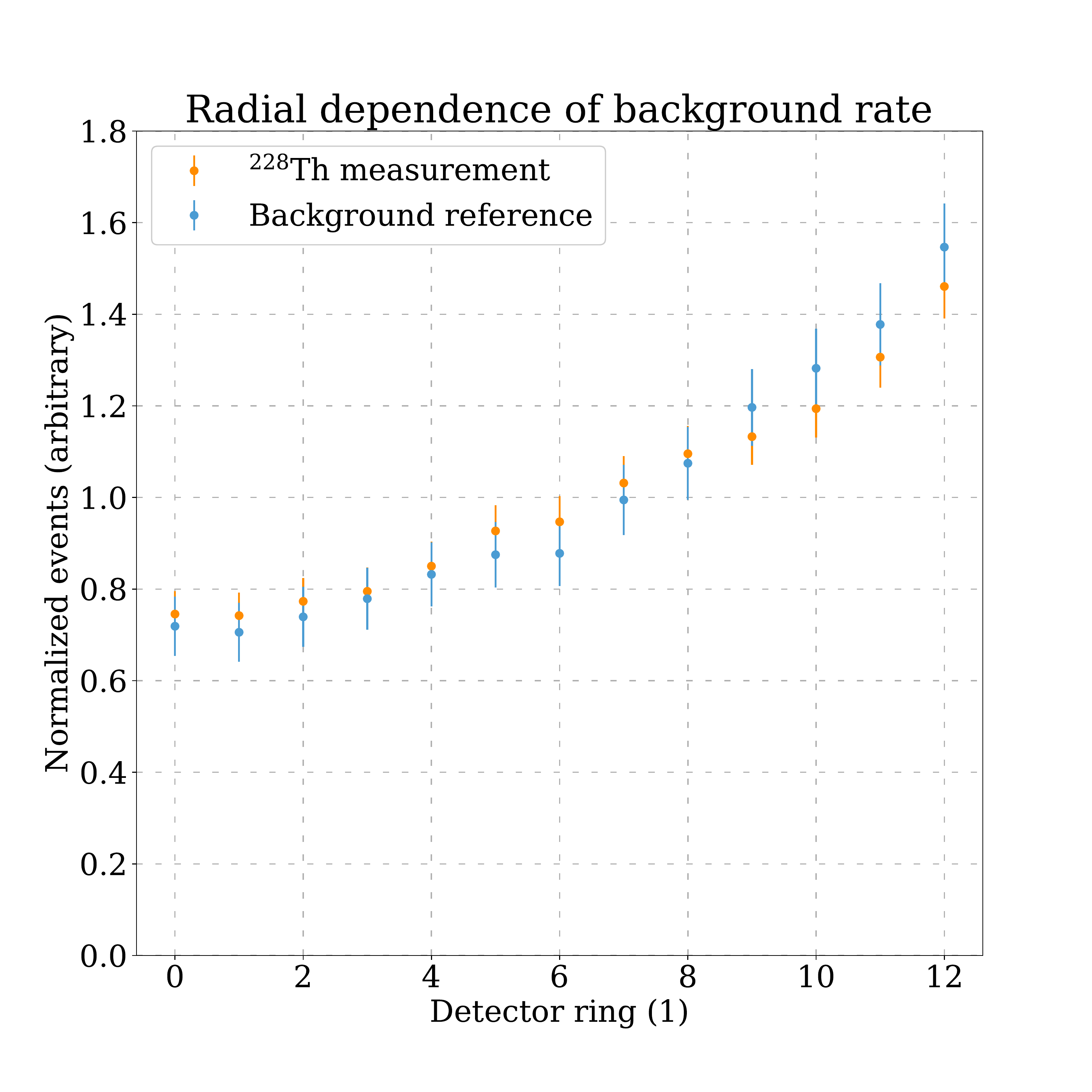}
	\caption{Normalized distributions of events for each detector ring. The distribution was normalized by dividing by the sum of all events and multiplying by the number of detector rings.}
	\label{fig:RadialEvents228Th}
\end{figure}

The background rate of the volume configuration is shown in figure \ref{fig:BGafter228ThVolume}. The following equation is used to describe the data:

\begin{equation}
  r(t) =  r_0 e^{-\frac{t}{\tau_1}} + c,
  \label{eq:thoriumFitExp}
\end{equation}

with $r(t)$ the rate at the detector in cps as a function of time $t$, $r_0$ the initial rate, and $c$ a constant rate offset in cps (dominantly non-\lead{212} related backgrounds).

The measured decrease of the rate over time with a half-life of \SI{10.66 +- 0.16}{\hour} matches very well the half-life of \lead{212}. The rate in the surface configuration (see figure \ref{fig:BGafter228ThSurface}) shows a similar exponential decrease with a somewhat longer half-life of \SI{10.91 +- 0.07}{\hour}. The model was extended for this fit from equation \ref{eq:thoriumFitExp} in order to model the decay chain including \bismuth{212}:

\begin{equation}
  r(t) =  \frac{N_2}{\tau_2} e^{-\frac{t}{\tau_2}} - N_1 \frac{\frac{1}{\tau_1 \tau_2}}{\frac{1}{\tau_2} - \frac{1}{\tau_1}} (e^{-\frac{t}{\tau_2}} - e^{-\frac{1}{\tau_1}}) + c,
  \label{eq:thoriumFitBateman}
\end{equation}

where $r(t)$ is the rate at the detector in cps as a function of time $t$, $N_1$ the initial number of \lead{212} atoms, $N_2$ the initial number of \bismuth{212} atoms, $\tau_1$ the life-time of \lead{212} in hours, $\tau_2$ the life-time of \bismuth{212} in hours, and $c$ a constant rate offset in cps (non-\lead{212} related backgrounds).

\begin{figure}
	\centering
	\includegraphics[width=0.5\columnwidth]{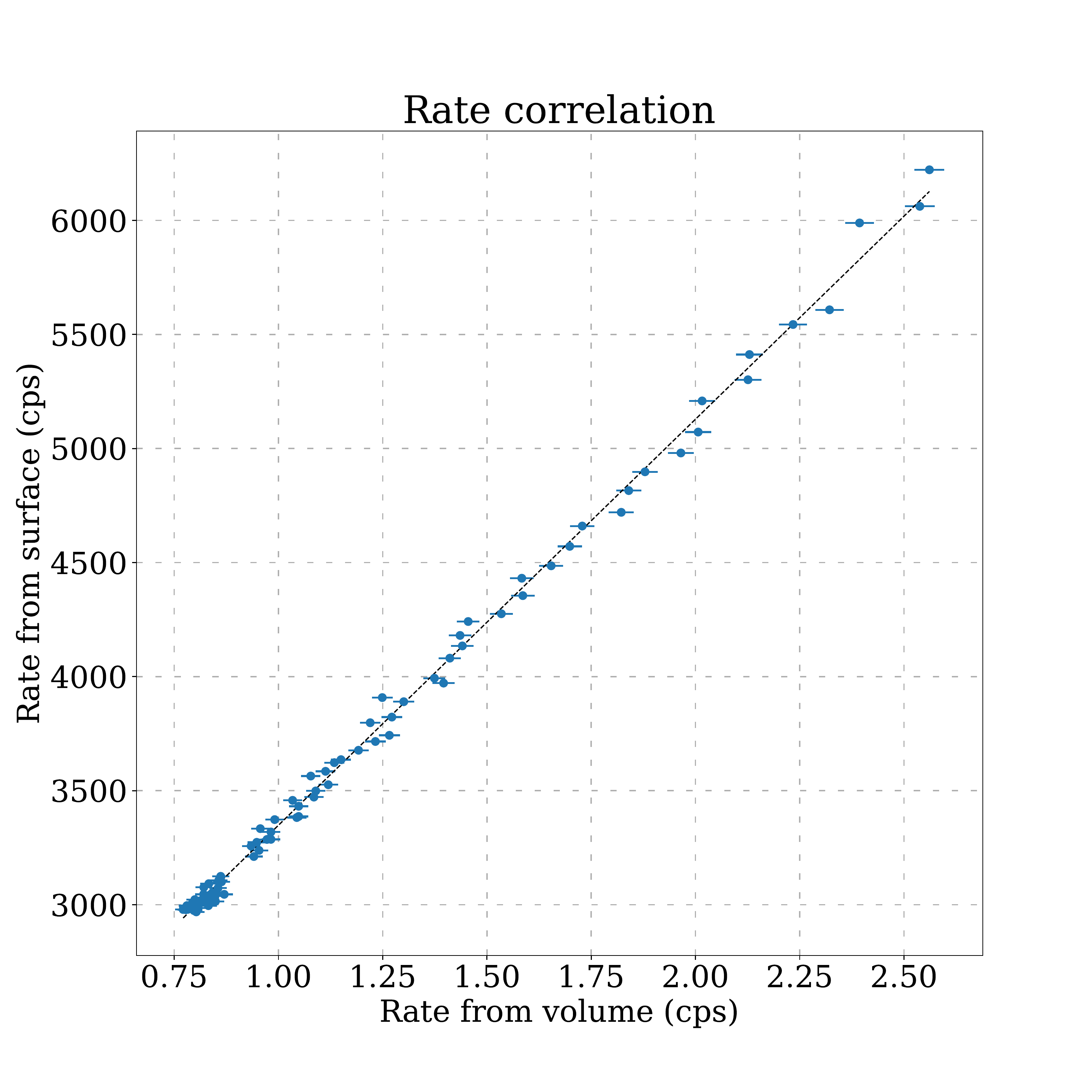}
	\caption{Correlation between the background rate from the spectrometer surface and the spectrometer volume. The Pearson correlation coefficient is 0.999.}
	\label{fig:VolumeSurfaceCorrelation}
\end{figure}

The measured half-lives of both measurement configurations agree very well and there is a very strong correlation between the rates in volume and surface configuration (see figure \ref{fig:VolumeSurfaceCorrelation}) with a Pearson correlation coefficient \cite{rodgers} of 0.999. The correlation was calculated by pairing the data points of a volume measurement with the subsequent surface measurement.  

The KATRIN detector is segmented into 148 pixels of equal area \cite{Amsbaugh2015}. The pixels are grouped into 13 concentric rings (4 pixels for the center ring, 12 pixels each for all other rings). The ring index increases with the radius, starting at zero for the ring in the center. Figure \ref{fig:RadialEvents228Th} shows the normalized events of each detector ring for a reference measurement, which was performed before the \thorium{228} exposure, and the first ten runs of the measurement in the volume field configuration. The similarity of both distributions points to the same underlying background mechanism.

The behavior of the background after the exposure of the Main Spectrometer matches very well to the expectation from the background assumption described in section \ref{sec:rydberg}. The subsequent $\alpha$-decays of the implanted \lead{212} on the inner spectrometer surface created a large number of secondary electrons which were observed in the surface magnetic field configuration. At the same time Rydberg atoms were produced in sputtering processes. As electrically neutral particles they propagated into the spectrometer volume where they could be ionized by the thermal radiation from the spectrometer surface.
The created electrons were accelerated towards the detector and create the observed background in the volume field configuration.
The whole process is time-wise dominated by the long half-life of \lead{212}.

\subsection{Radium source}

With the \radium{223} source we investigated in October 2018 the induced Main Spectrometer surface activity, and how it could generate background in the spectrometer volume. We also studied the characteristics of such a background.

After insertion of the \radium{223} source into the Main Spectrometer, the surface magnetic field configuration (see \cref{fig:fieldconfig}) was used to monitor the background evolution on the spectrometer surface. 
The observed behavior indicates accumulation of activity on the surface and can be described by the Bateman equation \cite{Bateman1910}  

\begin{align}
	A_{\text{Pb}}(t)=A_{\text{Ra}}(t_{0}) \frac{\lambda_{\text{Pb}}}{\lambda_{\text{Pb}}-\lambda_{\text{Ra}}}(e^{-\lambda_{\text{Ra}}t}-e^{\lambda_{\text{Pb}}t}) + A_{\text{Pb}}(t_{0})e^{-\lambda_{\text{Pb}}t}
	\label{equ:bateman}
\end{align} 

with $A_{\text{Ra}}(t_{0})$ and $A_{\text{Pb}}(t_{0})$ describing the activities of the isotopes before surface contamination was initiated at $t_{0}$ and $\lambda_{\text{Pb}}=$~\SI[exponent-product=\cdot]{3.19e-4}{\per\second}, $\lambda_{\text{Ra}}=$~\SI[exponent-product=\cdot]{7.02e-7}{\per\second} the decay rates of \lead{211} and \radium{223} \cite{Kirby1965, AitkenSmith2016}, respectively. 
The initial activity of lead in the walls is assumed to be negligible such that $A_{\text{Pb}}(t_{0})=$~0. 
The other \radium{223} daughters, \textsuperscript{219}Rn and \textsuperscript{215}Po, are negligible since their half-lives are only \SI{3.96}{\second} and \SI{1.78}{\milli\second}, leading to the saturation of activity in the spectrometer walls before the start of the measurement. 
A fit of \cref{equ:bateman} to the background rate observed after inserting the source, with the free parameter $A_{\text{Ra}}(t_{0})$ exhibits excellent agreement, confirming the spectrometer contamination with the \radium{223} daughter \lead{211}. 
After about four half-lives of \lead{211}, $\approx$\,\SI{2.5}{\hour}, the accumulated surface activity is close to saturation. 

\begin{figure}
	\centering
	\subfigure[~spectrometer surface]{%
		\includegraphics[width=0.5\columnwidth]{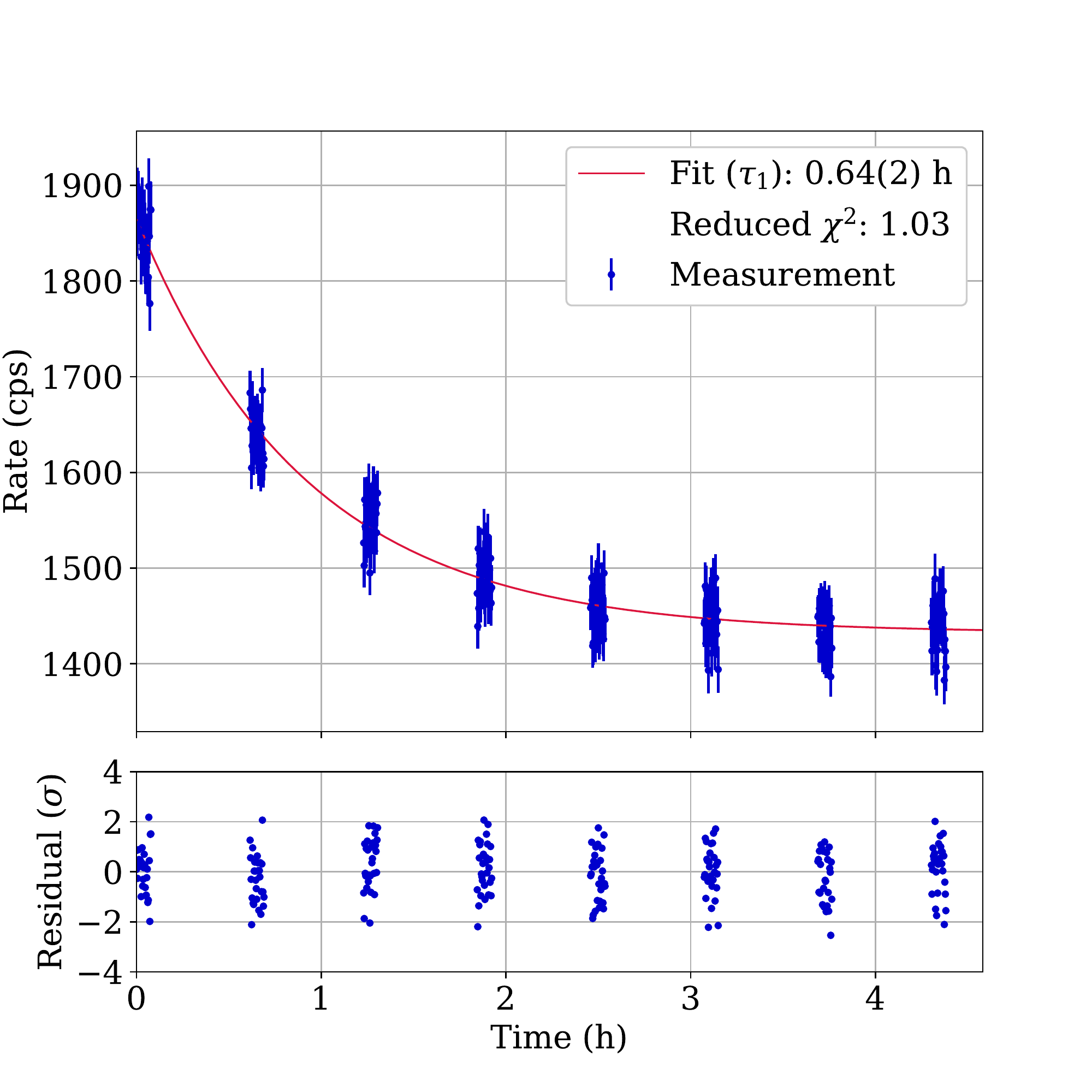}
		\label{fig:ra_surface}
	}%
	\subfigure[~spectrometer volume]{%
		\includegraphics[width=0.5\columnwidth]{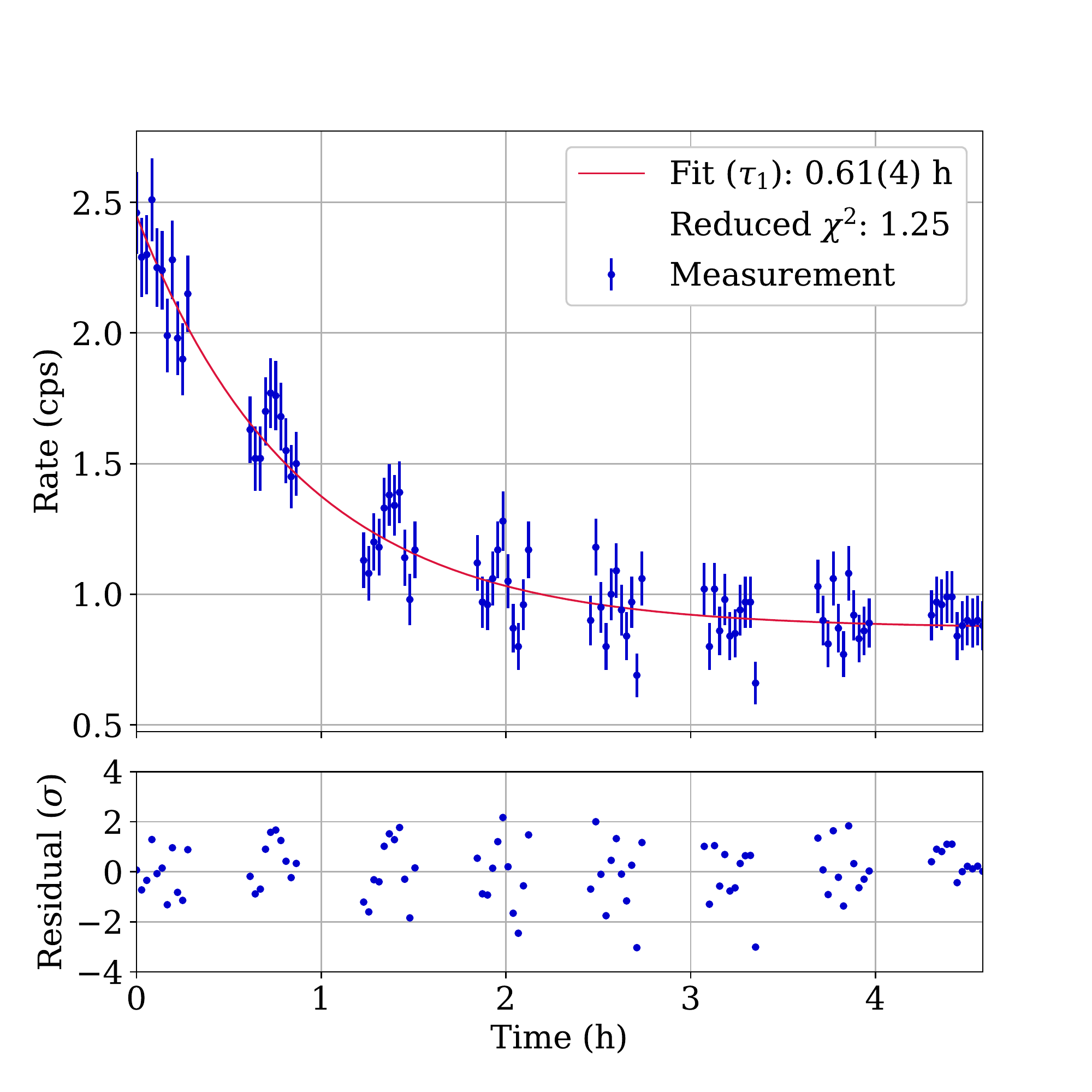}
		\label{fig:ra_volume}
	}%
	\caption{The measured background rates after the withdrawal of the \radium{223} source in the \protect\subref{fig:ra_surface} surface and \protect\subref{fig:ra_volume} volume magnetic field configurations, see \cref{fig:fieldconfig}. Both decay exponentially. A fit of \cref{eq:thoriumFitExp} to the measurement data is performed. The fitted half-life is in agreement with the expected one of \lead{211}, $t_{1/2}=$~\SI{36}{\minute}\,=\,\SI{0.6}{\hour}. The fitted constant offsets are \SI{1432 \pm 3}{\cps} and \SI{0.87 \pm 0.02}{\cps} for the surface and volume magnetic field configuration, respectively.}
	\label{fig:ra}
\end{figure}

Given a maximal surface activity, the $\alpha$-induced background in the volume, which is the relevant quantity for nominal KATRIN operation, was studied with the radioactive source retracted and the valve closed. 
By measuring alternately in the surface and volume magnetic field configurations, see \cref{fig:fieldconfig}, the backgrounds from the volume and the surface were studied. 
Each sequence in the surface magnetic field configuration lasted \SI{5}{\minute} and was followed by \SI{20}{\minute} in the volume configuration.
The corresponding background rates are shown in \cref{fig:ra}. 
Due to electrons created from radioactive decays, the background rates in spectrometer surface magnetic field configuration are not distributed according to a Poisson distribution.
In the spectrometer surface and volume magnetic field configurations, a decay half-life of \SI{0.64 \pm 0.02}{\hour} and \SI{0.61 \pm 0.04}{\hour}, respectively, is observed. 
Both are in agreement with the half-life \SI{0.60}{\hour} of \lead{211} \cite{AitkenSmith2016}, which is the dominant time-constant in the decay chain of \radium{223}.
The matching half-lives indicate the contaminated spectrometer surface as the origin of the background events in the volume, confirming observations with the thorium source. 
In addition, the data points are correlated with a Pearson correlation coefficient of 0.99\footnote{The correlation coefficient was calculated by binning both measurements into eight points and assuming that they were recorded at the same time}. 

\begin{figure}
	\centering
	\includegraphics[width=0.5\columnwidth]{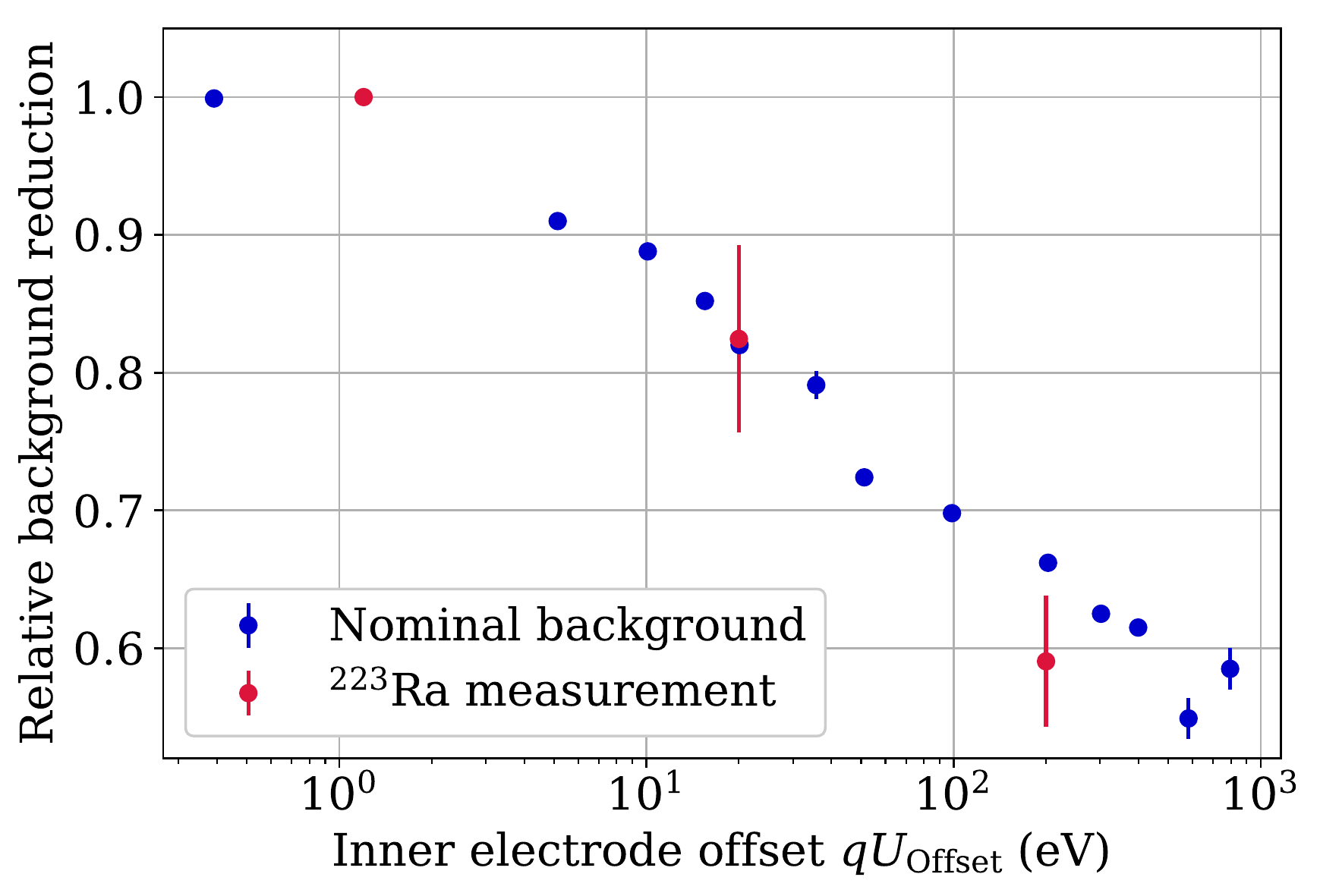}
	\caption{Background rate relative to zero inner electrode potential (the actual voltage reading slightly deviates for the zero setpoints) for the normal Main Spectrometer background and the $\alpha$ source induced background. }
	\label{fig:ie}
\end{figure}

To investigate the characteristics of the artificially created $\alpha$-source induced background, the measurement shown in \cref{fig:ra_volume} was repeated. 
This time the inner electrode voltage dependency of the \radium{223} induced background in the Main Spectrometer volume was studied in the first hour of the exponential decay by measuring at three different voltages, \SIlist[list-units=single]{0; -20; -200}{\volt}.
At each voltage we measured for about \SI{18}{\minute} so that the overall measurement time stayed within a range of significant \radium{223} background contributions.
To extract the voltage dependency from such a measurement, the measurement must be corrected for the exponential decay. 
The expected number of events was calculated under the assumption of an exponential decay at a constant voltage of $U_{\mathrm{IE}}=$~\SI{0}{\volt}, see \cref{tab:iedep} column 2. 
From the ratio of the expected number of counts to the observed number at different voltages, listed in column 3 of \cref{tab:iedep}, the dependency is obtained. 
The results are shown in \cref{fig:ie}. 
For different inner electrode offset potentials, the relative background reduction with respect to $U_{\mathrm{IE}}\approx$~0 is shown for the nominal Main Spectrometer \cite{Trost2019_1000090450} and \radium{223}-induced backgrounds. Both show a background reduction as $U_{\mathrm{IE}}$ decreases, which agrees well within uncertainties. 
This points to $\alpha$-decay induced background as the dominant Main Spectrometer background source. 
In the hypothesis of the Rydberg states as the background source, this voltage dependency can be explained by field ionization. 
With more negative inner electrode voltage more Rydberg states are ionized in the high electric field between spectrometer walls and wire electrodes and do not contribute to the background in the spectrometer volume. 
Since charged particles, like electrons, created at the spectrometer surface cannot penetrate the sensitive flux volume due to electric and magnetic shielding, our observations also indicate the existence of a neutral mediator, e.g. the Rydberg atoms. 

\sisetup{table-number-alignment=center}
\begin{table}
	\centering
	\begin{tabular}{l|
					S[separate-uncertainty, table-figures-uncertainty=1]|
					S[separate-uncertainty, table-figures-uncertainty=1]|
					S[separate-uncertainty, table-figures-uncertainty=1]}

		{$U_{\mathrm{IE}}$ (V)} & {\shortstack{observed {\#events}\\(nominal background subtracted)}}	& {\shortstack{expected \#events \\assuming $U_{\mathrm{IE}}=$~\SI{0}{\volt}}} &  {\shortstack{background relative to\\$U_{\mathrm{IE}}=$~\SI{0}{\volt}}}\\ 
		\hline
		-200						& 1079 \pm 45										& 1827\pm 126 &	 	0.59 \pm 0.05 \\
		-20						& 969 \pm 49										& 1176\pm 77 &	 	0.82 \pm 0.07 \\
		0						& 755 \pm 49										& 755\pm 49  &     1.00
	\end{tabular}
	\caption{Observed background events with the radium source at different inner electrode voltages after subtraction of the nominal spectrometer background. Given the exponential decay, the expected number of events is calculated assuming a constant inner electrode voltage of $U_{\mathrm{IE}}=$~\SI{0}{\volt}. From the ratio of expected and observed events the background rate relative to the rate measured at $U_{\mathrm{IE}}=$~\SI{0}{\volt} is calculated.}
	\label{tab:iedep}
\end{table}

\section{Conclusion}
\label{sec:conclusion}

We performed measurements with two $\alpha$-sources at the KATRIN Main Spectrometer to investigate whether an increased $\alpha$ activity on its surfaces could increase the electron background in its volume.  
Our results indicate $\alpha$-decays in the spectrometer walls as potential triggers of the background-creating process. 
Further, we show that such background exhibits the same radial distribution and inner electrode voltage dependency as the nominal Main Spectrometer background.
This points towards a significant contribution of $\alpha$-decay induced background to the residual background in KATRIN.
Given these findings, a neutral mediator that generates the background electrons in the spectrometer volume is required. 
A promising candidate is Rydberg atoms created from $\alpha$-decays in the spectrometer walls by sputtering processes. 
Low-energy electrons are produced through ionization of those highly excited atoms in the magnetic flux volume. 
The energy of photons from the thermal radiation of the walls at room temperature is sufficient to ionize some of the Rydberg atoms.
One possible countermeasure for such a background is a shift of the potential maximum towards the detector side, a so-called shifted analyzing plane \cite{proposalSAP}. 
Since Rydberg induced electrons are expected to be low-energy, only those generated in the volume between the maximal filter potential and the detector contribute to the detected background. 
By a spatial shift of the maximum potential, the relevant volume for background creation is reduced. 
First demonstration measurements have shown that this approach can reduce the background by more than a factor of two \cite{Pollithy2019, Lokhov2019}.

\begin{acknowledgements}
	We acknowledge the support of the Helmholtz Association (HGF), Ministry for Education and Research BMBF (05A17PM3, 05A17PX3, 05A17VK2, and 05A17WO3), Helmholtz Alliance for Astroparticle Physics (HAP), and Helmholtz Young Investigator Group (VH-NG-1055) in Germany; Ministry of Education, Youth and Sport (CANAM-LM2011019), cooperation with the JINR Dubna (3+3 grants) 2017–2019 in the Czech Republic; and the Department of Energy through grants DE-FG02-97ER41020, DE-FG02-94ER40818, DE-SC0004036, DE-FG02-97ER41033, DE-FG02-97ER41041, DE-AC02-05CH11231, and DE-SC0011091 in the United States.
	
	We acknowledge the support of the ISOLDE Collaboration and technical teams. 
	
	We acknowledge F. Müller and Y. Steinhauser for constructing the \radium{223} source module and R. Lang from Purdue University for lending us the thorium source. 
	
	JK acknowledges support by a Wolfgang Gentner Ph.D. Scholarship of the BMBF (grant no. 05E15CHA).
\end{acknowledgements}

\bibliography{literature}

\end{document}